\documentclass{article}
\usepackage{spconf,amsmath,graphicx}
\usepackage{tabularx}
{

\title{Multi-channel acoustic modeling using mixed bitrate opus compression}
\name{Aparna Khare, Shiva Sundaram, Minhua Wu}
\address{Amazon.com, Sunnyvale, CA}
\begin{document}
%
\maketitle
\begin{abstract}
Recent literature has shown that a learned front end with multi-channel audio input can outperform traditional beamforming algorithms for automatic speech recognition (ASR). In this paper, we present our study on multi-channel acoustic modeling using OPUS compression with different bitrates for the different channels. We analyze the degradation in word error rate (WER) as a function of the audio encoding bitrate and show that the  WER degrades by 12.6\% relative with 16kpbs as compared to uncompressed audio. We show that its always preferable to have a multi-channel audio input over a single channel audio input given limited bandwidth.  Our results show that for the best WER, when one of the two channels can be encoded with a bitrate higher than 32kbps, its optimal to encode the other channel with the highest bitrate possible. For bitrates lower than that, its preferable to distribute the bitrate equally between the two channels. We further show that by training the acoustic model on mixed bitrate input, upto 50\% of the degradation can be recovered using a single model. 
\end{abstract}
\section{Introduction}
\label{sec:intro}

Multi-channel modeling has become a popular alternative to conventional beamforming in recent literature \cite{kumatani2019multi}\cite{sainath2017multichannel}. This method of front-end processing opens up avenues for processing audio not just from microphones from a single microphone array but from microphones from distinct devices. With the popularity of voice enabled devices in the consumer markets, this opens up an interesting area of research. Different devices from different manufacturers, however, may not adhere to the same encoding standards. In addition, the bandwidth available for audio upload at different times might be different. Motivated by these questions, we wanted to study how multi-channel acoustic models perform when different audio inputs have different bitrates.

Prior work in this area has focused at developing compression algorithms to minimally impact speech recognition systems \cite{chazan2000low} or at analyzing the quality of the opus codec as it pertains to voice quality \cite{orosz2013performance} . In \cite{siegert2016measuring}, the authors evaluate the performance of different audio codecs on the frequency spectrum.  There is also prior work in ASR that examines mixed-bandwidth models that deal with data with different sampling rates \cite{mac2019large}. There has been some related work in the emotion recognition domain; the authors in \cite{garcia2015automatic} and \cite{albahri2016effect} show how different codecs and different bitrates affect emotion recognition accuracy. The authors in \cite{siegert2016emotion} analyze human emotion intelligibility as a function of the bitrates. The other related work is in the domain of multi-channel acoustic modeling where the traditional front-end processing algorithms like beamforming are learned within the acoustic modeling network. \cite{minhua2019frequency} and \cite{sainath2017multichannel} demonstrate that a data driven beamformer out-performs a traditional beamforming approaches. 

Our contribution in this paper is to study how the ASR model performance varies with the different bitrate encoding of the OPUS codec \cite{valin2012definition}.  We chose the OPUS codec for our analysis because it is an open source, industry leading standard for audio coding for multiple applications \cite{raju2015comparison}. The main benefit of OPUS over other codes is the hybrid approach, where it uses SILK for encoding information below 8khz and CELT for information above 8khz \cite{valin2012definition}, thus providing high quality for both speech and audio/music components. We analyze the performance of both the multi-channel and single channel acoustic models as a function of the audio input bitrate. We demonstrate that by training with mixed-bitrates we can recover some of the performance loss. Motivated by the multiple device use case described above, we also study the performance of the system if the multiple audio input channels to the network are not encoded with the same bitrate. The closest work to this was in \cite{narayanan2018toward}, where the authors show how the ASR WER varies with different codecs at a fixed bitrate encoding.

This paper is organized as follows. In Section \ref{sec:system}, we describe our model architecture, training techniques and the training and evaluation data. In Section \ref{sec:baselinebitrates}, we present our analysis of the ASR system performance with different bitrate audio inputs.  We discuss our experiments on how to optimally distribute bandwidth between multiple channels in Section \ref{sec:baselinemultibitrate}. In Section \ref{sec:mixtraining}, we describe the mixed-bitrate model training focused at recovering from degradation introduced by low bitrate encoding. Finally in Section \ref{sec:conclusion}, we present analyze and conclude our work, and present future work.

\section{System description}
\label{sec:system}

\begin{figure}[htb]
  \label{fig:res}
\begin{minipage}[b]{1.0\linewidth}
  \centering
  \centerline{\includegraphics[width=8.5cm]{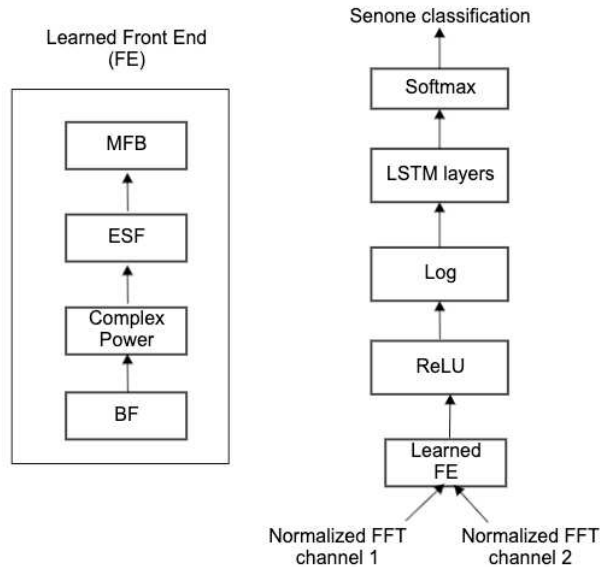}}
  \caption{System architecture}
\end{minipage}
\end{figure}

Our network architecture is similar to the Elastic Spatial Filtering system described in \cite{minhua2019frequency} and is shown in Figure \ref{fig:res}. The network takes in the FFT from a multi-channel audio input.  The first layer of the network is a block affine transform initialized by the coefficients of a super-directive beamformer that uses the spherically isotropic noise field \cite{himawan2010clustered}\cite{doclo2007superdirective}. This is followed by component that  computes the power of each of the frequency components in each of the look directions. This is followed by an affine transform that performs elastic spatial filtering as described in \cite{minhua2019frequency}. This block is followed by another affine transform that is initialized by mel-filter bank (MFB) coefficients, followed by a ReLU non-linearity and log to emulate computing log filter-bank energies. The ReLU component is used to ensure positive value input to the log component. These components form the learned feature extraction layers. This is followed by LSTM layers that classify the input to senones. The LSTM weights are initialized using a uniform distribution.

For all the experiments in the paper, we use a 2-channel speech input and a 256-point FFT without the Nyquist frequency and direct components. These features are mean and variance normalized and are then used as input into the network. The block affine transform is initialized with 12 different look directions. The output of the MFB layer is 64-dimensional. For the classification layers we use a 5-layer LSTM with 
768 cells in each layer. Our system uses 3183 senones.   We use an effective frame size of 30ms; the FFT is computed from a 10ms frame, and the beamformer, elastic spatial filtering and MFB affine transform layer operate at a frame size of 10ms. After the MFB layer, the features for 3 frames are concatenated and the rest of the processing consumes a 30ms frame.  Our training data consists of 620 hours of far-field data from our in-house dataset. The data is sampled at 16kHz. The array geometry used here is an equispaced six-channel microphone circular array with a diameter of approximately 72 milli-meters and one microphone at the center. For all our experiments, we use 2 microphones from the opposite sides of the microphone array. All the models are trained with the cross-entropy objective followed by the sMBR objective function \cite{vesely2013sequence}. Our test data consists of 33,000 far-field  speech utterances, and our development set consists of 17,000 similar utterances.

\section{Performance as a function of encoding bitrates}
\label{sec:baselinebitrates}
For this study, we train our baseline multi-channel model with uncompressed audio as input. The architecture and training data are described in Section \ref{sec:system}.  We use the development set to tune the acoustic scale for our system.  For this specific study, our development set is also uncompressed audio. We use the OPUS encoder \cite{vos2013voice} to encode the uncompressed audio in our test set using different bitrates using the constant bitrate setting \footnote{Our experiments didn't show a significant difference between the variable bitrate and constant bitrate OPUS settings}. The degradation at lower bitrates can be explained by the distortion introduced in the speech characteristics by the codec and removal of useful speech information. 

\begin{figure}
  \label{fig:spcgram}
\begin{minipage}[b]{1.0\linewidth}
  \centering
  \centerline{\includegraphics[width=8.5cm]{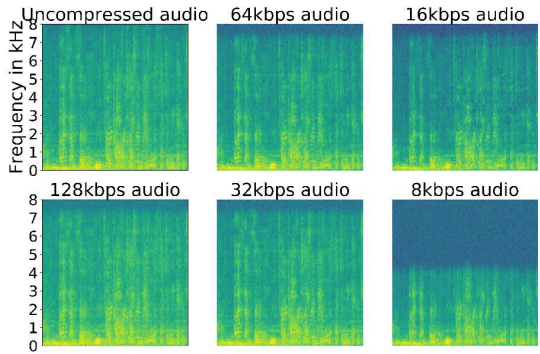}}
\caption{Spectrogram for one test utterance at various bitrates}
\end{minipage}
\end{figure}

The WER with the uncompressed test data is our baseline; for the analysis we use the audio from the same test set and encode it with 8kbps, 16kbps, 32kbps, and 128kbps per channel. Table \ref{table:tableno1} shows the relative word error rate (WER) degradation with the different bit rate audio inputs for the multi-channel model. As our results show, the WER degrades sharply after 16kbps, this is expected as the performance with the 8kbps encoding degrades drastically since OPUS encodes the audio as narrowband at 8kbps. With a bitrate of 128kbps per channel, the degradation reduces to 2.4\% relative, demonstrating that most of the speech characteristics required for ASR are preserved in this case. We also visualize the spectrogram of a single audio example with various bitrates in Figure \ref{fig:spcgram}. As it can be seen from the figure, 8kbps encoding has no information above 4khz.  The information loss, specially at higher frequencies is clear in both the 16kbps and 128kbps encoding.

\begin{table}[htb]
  \caption{Relative WER degradation with different bitrates for multi-channel input}
  \label{table:tableno1}
  \centering
  \begin{tabular}{|c|c|}
  \hline
    \textbf{Bitrate per channel}    & \textbf{relative WER degradation \%}          \\
    \hline
   Uncompressed     & -    \\
        \hline
    8 &  202.1 \\
        \hline
    16    & 12.6   \\
        \hline
    32	 & 6.6  \\
        \hline
    128 &  2.4\\
        \hline
      \end{tabular}
\end{table}

  \begin{table*}
\centering
  \caption{Relative WER degradation with mixed bitrates input}
  \label{table:tableno4}
  \begin{tabular}{|c|c|c|c|c|c|}
   \hline
	\textbf{Total bitrate} & \textbf{Bitrate for channel 1}   & \textbf{Bitrate for channel 2} & \multicolumn{3}{|c|}{\textbf{Relative WER degradation}} \\
	\hline
	& & & Overall & Clean & Noisy \\
	\hline
	Uncompressed & Uncompressed & Uncompressed & - & - & - \\
	\hline
      264 & 256 & 8 & 11.1 &  10.5 & 10.9  \\
      264 &132 & 132 & 2.7  & 3.7  & 2.2\\
    \hline
       252 & 256 &  16 & 4.1  & 3.7  & 4.4  \\
      252 & 136 & 136 & 2.7 & 3.7  & 2.2\\
    \hline
       288 & 256 &  32 & 1.5  & 1.9 & 1.1 \\
      288 & 144 & 144 & 2.7 & 3.7  & 2.2\\
    \hline
      320 & 256 & 64 & 0  & 0 & 0\\
      320 & 160 & 160 & 2.8 & 4.3  & 2.2 \\
    \hline
      384 & 256 & 128 & 0  & 0 & 0 \\
      384 & 192 & 192 &  2.4 &  3.7  & 2.2\\
    \hline
  \end{tabular}
  \end{table*}

\begin{figure}[t]
  \label{fig:res1}
\begin{minipage}[b]{1.0\linewidth}
  \centering
  \centerline{\includegraphics[width=8.5cm]{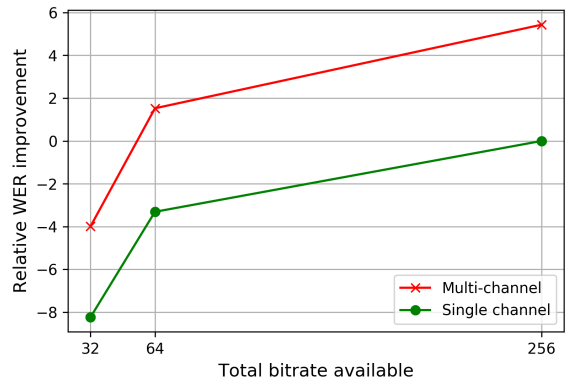}}
\caption{Relative WER improvement as a function of bitrates, positive values indicate improvement in WER}
\end{minipage}
\end{figure}

We compare this degradation with similar inputs to a single channel model in Figure \ref{fig:res1}. For a fair comparison, we compare the WER for the single channel with bitrate $2x$ with the multi-channel model with bitrate  $x$ per channel, since the number of available bits for encoding the data is equal in those cases. The single channel model does not have the beamforming and the elastic spatial filtering layers. The rest of the components are initialized and trained similar to the multi-channel model . For this analysis, the baseline is the single channel model trained with uncompressed audio as input. The multi-channel model outperforms the single channel model for all bitrates by 4-6\% relative depending on the total bits available for encoding. Comparing the performance at 256kbps for the single channel model and 64kbps for the multi-channel model, we note that the multi-channel model outperforms the single channel model by 1.52\% relative. From these results, we infer that the OPUS compression preserves sufficient phase information from the waveforms to improve performance using the learned beamformer.  The take away from this study is that given limited bandwidth, we should always prefer dividing the bandwidth into multiple channels instead of using all of it for encoding a single channel. This study demonstrates that we can potentially leverage from the multi-channel model architecture when the inputs are from different devices with different , in which case performing beamforming on a single device is not practical. 

\section{Distributing limited bandwidth among multiple channels}
\label{sec:baselinemultibitrate}

In Section \ref{sec:baselinebitrates}, we have established that we should distributed limited bandwidth to multiple channels instead of a single channel. For the second part of the study, we wanted to understand how to distribute limited  bandwidth between multiple channels.  For this experiment, we construct two different test sets. For the first set of experiments, we fix the input to one of the channels to be the uncompressed audio and vary the bitrate of the second channel to 8,16,32, 64 and 128kbps. The second set is constructed where the kilo-bytes per second available is uniformly distributed between the 2 channels. For the purpose of the experiment, we assume that the uncompressed audio is equivalent to audio with a 256kbps rate, since our experiments showed that audio compressed with 256kbps produced the same WER as uncompressed audio. The baseline for this experiment is uncompressed audio for both channels as input. Our results are shown in Table \ref{table:tableno4}. 

From the results, we can see that given limited bandwidth, it is preferable to have a single channel with higher bitrate than to divide the bandwidth equally, given the other channel has a bitrate higher than 16kbps. We further analyze these results by noisy and  clean conditions. We  define the noisy condition as the SNR $<$ 5dB and clean conditions as SNR $>=$ 5dB given the far-field nature of the audio. These results show that in both noisy and clean conditions, having a single channel encoded with the highest bit rate  is beneficial over distributing the bitrate between the two channels equally. This demonstrates that the learned beamformer performs better when there is at least one channel with no degradation rather than both channels having some degradation, even when the bitrate is as low as 32kpbs. In fact  we can recover performance when the bitrate of one channel is greater than 64kbps as long as we retain high bitrate on the other channel.  However when the bitrate for one channel gets as low as 8kbps, its preferable to distribute the bitrate between the two channels evenly. 

\section{Mixed bitrate training}
\label{sec:mixtraining}

Next, we wanted to explore if the degradation can be compensated by matching the training data with the test data. For this purpose, we did 2 different sets of experiments. The first was to train individual multi-channel models for each bit-rate encoding we analyzed to see if we can compensate for the degradation . The second experiment was focused towards understanding if a single acoustic model can work well for different bitrates for different channels. We trained a model with mixed bitrates; for each training example, we picked an encoding bitrate uniformly from the set $\{uncompressed,16,32,64,128\}$. In order to keep the comparison fair, we used  the same number of training examples for  training the mixed bitrate model and did not sample any training example multiple times with different bitrates. For each of these models, the acoustic scale was tuned on the development set that matched the training set.

Table \ref{table:tableno3} shows the results of these experiments. As expected, the training with matched bitrate encoding shows improvement over the baseline. For 32 kbps, the word error rate degradation reduces by 50\%. The degradation with 16kbps can be attributed to the loss of information in encoding that cannot be  recovered by matching the test data. The single mixed bitrate model can recover some of the degradation for all the different bitrate encodings we studied. For bitrates higher than 16, it is not as good as the matched training but better than the baseline. It is interesting to note however that for 16kbps the model can perform better than the matched training. We posit that this is because the matched model doesn't see critical information for higher frequencies during training and hence cannot recover the errors. We also note that with this single model, we see a 3.4\% degradation in WER when the audio input is uncompressed. This can be explained by the mismatch in the training and test data. This degradation can be recovered if the each training sample presented to the model with multiple encoding bitrates. 

\begin{table}
  \caption{WER degradation with different bitrate models}
  \label{table:tableno3}
  \centering
  \begin{tabular}{|c|c|c|}
  \hline
    \textbf{Test Bitrate per channel}    & \textbf{Training data}     & \textbf{WERR \%}     \\
    \hline
    Uncompressed & Uncompressed & - \\
    16     & Uncompressed  & 12.6  \\
    16     & 16 & 10.0 \\
    16     & mixed bitrates & 7.9 \\
        \hline
    Uncompressed & Uncompressed & - \\
    32     & Uncompressed  & 6.6   \\
    32     & 32 & 3.3 \\
    32     & mixed bitrates & 4.1 \\
        \hline
    Uncompressed & Uncompressed & - \\
    128     & Uncompressed  & 2.4  \\
    128   & 128 & 1.6 \\
    128    & mixed bitrates & 2.1 \\
        \hline
     Uncompressed & Uncompressed & - \\
    Uncompressed     & mixed bitrates  & 3.4  \\
        \hline
      \end{tabular}
\end{table}

\section{Analysis and Conclusions}
\label{sec:conclusion}
In this work, we studied the degradation of WER as a function of bitrates available for the input audio. We showed that even with low bitrate encoding, the multi-channel model can outperform the single channel model, which opens up multi-channel modeling with multiple devices that may not support the same bitrate. Prior work with multi-device acoustic modeling has not focused on the mixed bitrate input \cite{wang2019stream}\cite{sadhu2019m}. We also demonstrated that when limited bandwidth is available, we should always choose a multi-channel input over a single channel input. For the best performance, it is optimal to encode one channel with the highest bandwidth available given the second channel has sufficient bandwidth available. Finally we demonstrate that by training the model with variable bitrates, we can potentially use a single model for inputs with all bitrates which is important for practical applications.

For our future work, we would like to explore an attention based mechanism to replace the beamformer, similar to the work the authors have proposed in \cite{wang2019stream} . This can be especially useful with mixed bitrate inputs so that the model can leverage the best information required from both channels for the task. In the future, we would also like to extend our study to using real data from multiple devices.

We would like to acknowledge the support of our colleague, Kenichi Kumatani, for this work.

\bibliographystyle{IEEEbib}
\bibliography{refs}

\end{document}